\documentclass[aip, preprint, pop]{revtex4-1}
\usepackage{times}		
\usepackage{booktabs}
\usepackage{placeins}

\usepackage{amssymb,amstext,amsmath, epsf}
\usepackage{graphicx,graphics,epsfig,wrapfig}
\usepackage{listings}
\usepackage{float}
\usepackage{esint}

\usepackage{color}
\usepackage[english]{babel}
\usepackage[T1]{fontenc}
\usepackage[utf8]{inputenc}

\def\apj{{ApJ}}
\def\apjl{{ApJL}}

\def\nat{{Nature Physics}}
\def\prl{{PRL}}

\def\pre{{PRE}}

\def\apss{{Astrophysics and Space Science}}

\begin{document}

\title{Radiation From Particles Moving in Small-Scale Magnetic Fields Created in Solid-Density Laser-Plasma Laboratory Experiments}

\author{Brett D. Keenan}
\email{bdkeenan@ku.edu}
\affiliation{Department of Physics and Astronomy, University of Kansas, Lawrence, KS 66045}
\author{Mikhail V. Medvedev}
\affiliation{Department of Physics and Astronomy, University of Kansas, Lawrence, KS 66045}


\begin{abstract}
Plasmas created by high-intensity lasers are often subject to the formation of kinetic-streaming instabilities, such as the Weibel instability, which lead to the spontaneous generation of high-amplitude, tangled magnetic fields. These fields typically exist on small spatial scales, i.e.\ ``sub-Larmor scales''. Radiation from charged particles moving through small-scale electromagnetic (EM) turbulence has spectral characteristics distinct from both synchrotron and cyclotron radiation, and it carries valuable information on the statistical properties of the EM field structure and evolution. Consequently, this radiation from laser-produced plasmas may offer insight into the underlying electromagnetic turbulence. Here we investigate the prospects for, and demonstrate the feasibility of, such direct radiative diagnostics for mildly relativistic, solid-density laser plasmas produced in lab experiments.  
\end{abstract}

\maketitle

\section{Introduction}
\label{s:intro}

For over a decade, the production of strong ($>$ megaGauss) magnetic fields/turbulence in solid-density plasmas, generated by the irradiation of a target with high-intensity lasers, has been observed in a diverse set of laboratory experiments \citep{sandhu, sarri, wagner, gopal, mondal12}. Understanding and controlling electromagnetic turbulence in these environments is critical to studies in the fusion energy sciences, and for the inertial confinement concept \citep{ren04, tatarakis03}, in particular. Additionally, electromagnetic turbulence is a crucial aspect of numerous astrophysical systems such as gamma-ray bursts and supernova shocks \citep{medvedev09b, medvedev06, reynolds12, kamble+14}. 

These magnetic fields can be generated by a number of mechanisms -- e.g., by the misalignment in plasma temperature and density gradients (Biermann Battery), or by an induction field produced by the flux of fast electrons via the ponderomotive acceleration \citep{belyaev}. At relativistic intensities ($> 10^{18} \ W/cm^2$) and ultrashort pulse durations ($< 1 \ ps$), magnetic fields can also be generated via an electron-driven Weibel-like instability \citep{belyaev, sarri}. Unlike the pure Weibel instability driven by the plasma temperature anisotropy \citep{weibel59}, this Weibel-like instability is initiated by counterstreaming electron beams \citep{fried59} consisting of a ``hot'' beam (arising immediately following the target's interaction with the high-intensity laser) and a returning (shielding) ``cold'' electron beam. Initially, the net current is zero; however, the Weibel-like instability subsequently grows, leading to the formation of separated current filaments -- the source of a quasi-static magnetic field configuration. These Weibel fields reside on a ``small-scale'' -- since the spatial scale (i.e., the correlation length) is dictated by the electron skin-depth (which is typically less than, or similar to, the electron gyro-radius). The current filaments may further evolve, via coalescence/tearing/screw instabilities, into current channels \citep{medvedev+05, polomarov+08, shvets+09}, which further initiate filamentary magnetic structures \citep{mondal12}. 

Additionally, Weibel-like electromagnetic fields have been implicated in the mediation of astrophysical collisionless shocks in (initially) unmagnetized plasma media \citep{medvedev99, silva03, frederiksen+04, spitkovsky08, nishikawa+09, bret14}. It is strongly believed that presently existing laser facilities, such as OMEGA/OMEGA EP and NIF, will eventually observe these Weibel-mediated shocks in the laboratory, i.e., to make a ``gamma-ray burst in a lab'' \citep{medvedev07, medvedev08, medvedev+09}. In contrast to the aforementioned solid-density plasmas, these plasmas flow freely in-between laser ablated metal plates \citep{fox+13, sakawa13, bret14, huntington15, park+15}. This is achieved via weaker laser intensities and longer pulse durations ($\sim 10^{14} \ W/cm^2$ and $\sim 1 \ ns$, for a recent Omega laser experiment) -- although higher intensities are believed to be required for the creation of a shock \citep{sakawa13, huntington15}. Recently, the formation of filamentary structures indicative of ion-driven Weibel-like magnetic fields have been observed in a scaled laboratory experiment at the Omega Laser Facility \citep{fox+13, huntington15, park+15}. 

Electrons moving in small-scale magnetic turbulence emit radiation that is distinct from both synchrotron and cyclotron radiation. In the context of plasma astrophysics, this radiation is known as ``jitter'' radiation. However, to prevent confusion  with the ``jitter'' of electrons in the laser wave-field, we provisionally adopt a new term: ``Weibler'' radiation. We choose this term because ``jitter'' radiation is often associated with Weibel-like magnetic fields.

Thus, Weibler (``jitter'') radiation, via its spectrum, offers considerable information about the statistical properties of the underlying magnetic turbulence  \citep{medvedev00,medvedev06,medvedev11,RK10,TT11, keenan13}. We will show that the direct observation of mildly relativistic Weibler radiation may be feasible in the laboratory setting. We will focus our attention upon the experiment discussed in Ref. \citep{mondal12}. This experiment provides a concrete example of an applicable laser plasma. Additionally,  the Ref. \citep{mondal12} experiment constructed, directly from data, the magnetic (spatial) power spectrum -- an estimate of which is necessary to predict the Weibler radiation spectra. A considerable amount of what is explored here is applicable to (short duration) relativistic laser-plasma experiments, in general. 

The rest of the paper is organized as follows. Section \ref{s:mond} presents the details of the Ref. \citep{mondal12} experiment. In Section \ref{s:jitter}, we briefly review the Weibler radiation theory. Then, using some simple estimates, we examine the observability of Weibler radiation in the laboratory. In Section \ref{s:brems}, we explore competing radiation mechanisms -- particularly, thermal Bremssrahlung emission. We show that its contribution is negligible compared to the Weibler emission in a spectral window of interest. Section \ref{s:rad_cool} compares the radiative cooling times for both radiation mechanisms, showing that cooling is insignificant for the typical duration of an experiment. In Section \ref{s:pow_spec}, allowing for some simplifying assumptions, we predict the radiation spectrum to be observed in the experiment. Finally, Section \ref{s:concl} is the conclusions. We use cgs units throughout the paper.

\section{The Weibel Instability in Laser-Plasma Experiments}
\label{s:mond}

In the experiment discussed in Ref. \citep{mondal12}, conducted at the Tata Institute of Fundamental Research (TIFR), an aluminum coated, BK-7 glass target was irradiated by a $10^{18} \ W/cm^2$ ($800\ nm$, $30\ fs$ duration) laser pump beam -- thereby creating a plasma in the aluminum layer (with thickness several times larger than the electron skin-depth) of the target. A low-intensity probe beam ($400\ nm$, $80\ fs$) was then introduced at a delay to the initial pumb beam. This probe beam was then reflected by the corresponding critical plasma surface. By exploiting the Cotton–Mouton effect, the strength, spatial, and temporal evolution of the generated magnetic fields were inferred by measuring the ellipicity induced in the probe beam's polarization.

The observed magnetic fields were very intense, with a maximum value $\sim 63\ MG$. Additionally, the fields were relatively long-lived -- existing on a several picosecond time-scale -- which is about a hundred times longer than the laser duration time-scale. These fields initially grow on a femtosecond time-scale and on spatial scales comparable to the electron skin-depth at the critical surface, $d_e \equiv c/\omega_{pe} \sim 0.1\ \mu{m}$ -- which is smaller than the probe spatial resolution of a few microns; consequently, their initial development was not directly observable. Nonetheless, the Weibel fields further evolved via mechanisms such as Kelvin-Helmholtz (KH) like instabilities (driven by fluid-like velocity shears). Finally, the random magnetic filamentary structures eventually exist on a picosecond time-scale and on a many micron spatial-scale -- allowing their detection.

In Ref. \citep{mondal12}, it was reported that the spatial spectrum of the magnetic field (in the target's transverse plane) is well approximated by an inverse power-law which extends to spatial scales below the electron skin-depth. Furthermore, the spectral shape remains largely unchanged over a $\sim 10\ ps$ time-scale. This result was additionally confirmed by 2D Particle-in-Cell (PIC) simulations. The PIC simulations further indicated that the magnetic field development is largely insensitive to the initial electron ($10\ eV$) and ion ($1\ eV$) temperatures. The final PIC ion temperatures were in the range  $4-8 \ keV$. The final electron temperature ($300-600\ keV$; $t \sim 10\ ps$) implies that the electrons are mildly relativistic; i.e.\ $\gamma_e \equiv 1/\sqrt{1 - \beta^2} \sim 2$, where $\beta \equiv v/c$ is the normalized electron velocity, and $c$ is the speed of light.

It is worthwhile to note that the scale of the magnetic field is dictated by the electrons in these solid-density plasma experiments. In contrast, the Weibel instability in laser ablated plasma flows is mediated by the ions. Consequently, the spatial scale of these Weibel magnetic fields will be on the order of the ion skin-depth. For this reason, these magnetic fields will not be sub-Larmor-scale (``small-scale'') with respect to the electrons; thus, the electrons will not emit radiation in the small-deflection Weibler regime. Therefore, the magnetic fields are not so easily identifiable by the internal radiation production of the plasma electrons. Rather, proton-radiography or Thomson scattering, via the injection of external particles, is the prescribed diagnostic tool \citep{huntington15, sakawa13}. 

In principle, the sub-Larmor-scale ions should emit Weibler radiation, but this will be orders of magnitude less intense (because of their higher mass) than the radiation produced by electrons via alternative radiation mechanisms. In addition, plasma dispersion would certainly screen out any ion Weibler radiation, since the characteristic emission frequencies will be well below the electron plasma cutoff frequency. Thus, we do not anticipate that our results will be immediately applicable to the laser setups, such as NIF and OMEGA/OMEGA EP, as they stand currently. These experiments would, rather, likely require a modification of the setup to realize the creation of a solid-density-type plasma -- as explored here.

\section{Weibler radiation}
\label{s:jitter}

The question we address here is whether or not the plasma electrons emit \emph{Weibler radiation} in setups similar to the Ref. \citep{mondal12} plasma experiment. Furthermore, is this radiation directly observable in the framed experiment? Firstly, we must determine if the Weibler prescription is appropriate, given the experimental parameters. Three principal parameters determine the Weibler regime: the magnetic field strength, the electron velocities, and the magnetic field correlation length. The first two parameters are known scaling functions of the laser intensity, $I$ and wavelength, $\lambda$. For a given intensity and wavelength, the  (final) ``forward'' electron temperatures are, respectively \citep{hatchett}
\begin{equation}
k_B{T_e} \sim U_{pond} \sim 1 \ MeV \times \sqrt{\frac{I\lambda^2}{10^{19}}},
\label{scale_temp}
\end{equation} 
where $k_B$ is the Boltzmann constant, $I$ is in units of $W \ cm^{-2}$, $\lambda$ is in units of $\mu{m}$, and $U_{pond}$ is the ponderomotive potential of the incident laser beam. Substitution of the Ref. \citep{mondal12} parameters gives the electron temperature of $253 \ keV$. This is comparable to the PIC simulation (final) ``effective'' electron temperature $300-600 \ keV$.

The laser generated Weibel magnetic field is predicted to have the maximum value \citep{belyaev} 
\begin{equation}
B_{Weibel}^{max} \sim \frac{m_e\omega_{pe}c}{e},
\label{weibel_max}
\end{equation} 
which follows from the fact that the Weibel magnetic field is of the same order as the (circularly polarized) laser electric field \citep{krainov}. Eq. (\ref{weibel_max}) suggests $B^{max}_{Weibel} \approx 171 \ MG$ (for $d_e \sim 0.1 \ \mu{m}$), which is similar to the maximum experimental value of $\sim 100 \ MG$. 

Next, the correlation length of the magnetic field is indicated by the characteristic wave number of the turbulent spectrum, $k_{Bf}$. Given an inverse power-law spectrum for the magnetic fluctuations, $k_{Bf}$ is the minimum wave number, $k_{min}$. The small-scale Weibel magnetic fields exist on scales comparable to the electron skin-depth. This sets the correlation length of magnetic field; hence, $k_{min} \sim d_e^{-1}$. 

Now, electrons moving in a random, static, magnetic field ${\bf B}$ will produce radiation in the small-deflection Weibler regime if their Larmor radius $r_L$ is much smaller than the magnetic field correlation length $\lambda_B$ -- the ratio of which, we call the ``gyro-number'' \citep{keenan15}
\begin{equation}
r_L\lambda_B^{-1} \sim k_{Bf}{r_L} = k_{Bf}\frac{{\gamma_e}m_e\beta{c^2}}{e{\it B}} \equiv \rho,
\label{scale_para}
\end{equation} 
Where ${\it B}$ is an appropriate statistical average of the magnetic fluctuations. In the Ref. \citep{mondal12} experiment, the spatially/temporally averaged magnetic field ($\approx 100 \ MG$) was slightly larger than the maximum value of $63\ MG$. We have elected to take ${\it B} \sim {\it B}_{max} \approx 63\ MG$. 

Lastly, $\rho$ will necessarily be small in the initial stages of the electron acceleration. So, we consider only the final time velocities (obtained from the PIC simulations) which are $v/c \equiv \beta \approx 0.78-0.89$. Then, finally, considering $\beta_{min} \equiv 0.78$, the gyro-number, $\rho \approx 4$. Thus, since $\rho$ is slightly greater than unity, the radiation regime will be predominantly characteristic of the, mildly relativistic, Weibler regime. 

Nonetheless, the observability of the radiation is subject to a number of conditions. In the following subsections, we will outline and roughly estimate these limiting factors. Obviously, this list may not be exhaustive, but we will address the most apparent concerns. 

\subsection{The Weibler Frequency}

Is the Weibler radiation production time-scale small enough to temporally resolve the spectrum? This question may be answered by considering the characteristic Weibler frequency \citep{keenan15},
\begin{equation}
\omega_{jn} \equiv \gamma_e^2{k_{min}}\beta{c}.
\label{jitter_freq}
\end{equation} 
Considering only the final electron temperatures (i.e.\ the velocities $\beta = 0.78, 0.89$), the Weibler frequency is
\begin{equation}
\omega_{jn} \sim 6\times10^{15}-1\times10^{16} \ rad/s,
\label{jitter_freq_numbers}
\end{equation} 
indicating that the radiation is in the Extreme Ultraviolet (EUV) part of the EM spectrum. To avoid shielding by the plasma, $\omega_{jn}$ must be greater than the electron plasma frequency, $\omega_{pe} = c/d_e$. The electron plasma density, at the critical surface, is indicated by the skin-depth, $d_e \sim 0.1\ \mu{m}$. The corresponding plasma frequency is 
\begin{equation}
\omega_{pe} \sim 3\times10^{15} \ rad/s.
\label{plasma_freq}
\end{equation} 
Thus, the Weibler frequency is slightly larger than the critical plasma frequency. This indicates that plasma dispersion will play an important role in determining the spectral shape of the radiation. 

In contrast, non-relativistic electrons would emit cyclotron radiation in large-scale (i.e., weakly inhomogeneous or uniform) magnetic fields. In this case, the mean magnetic field (acting in place as an ambient, uniform field) will admit a slightly broadened cyclotron component due to mild relativistic effects. The cyclotron frequency is
\begin{equation}
\omega_{ce} = e\langle B \rangle/m_e c.
\label{synch_freq}
\end{equation} 
With $\langle B \rangle \sim 100\ MG$, this is roughly
\begin{equation}
\omega_{ce} \sim 2\times10^{15} \ rad/s,
\label{synch_freq_est}
\end{equation} 
This is slightly below the plasma cutoff, $\omega_{pe}$. Thus, this cyclotron feature may not be readily observable -- while, in contrast, the Weibler frequency will be larger by a factor of a few. Furthermore, The isotropic Weibler spectrum has a high-frequency break at
\begin{equation}
\omega_{bn} = \gamma_e^2k_{max}\beta{c},
\label{jitter_break}
\end{equation} 
where $k_{max}$ is the maximum turbulent wave number; i.e.,\ the inverse of the turbulent wavelength at the shortest spatial-scale. The Weibler and the break frequencies determine the window where most of the radiation is emitted, $\omega_{jn}\lesssim\omega\lesssim\omega_{bn}$.

Next,  in order to well-resolve the radiation spectrum, one must observe the signal over several characteristic time-scales. Given a mildly relativistic electron, this time-scale must be several $\omega_{jn}^{-1}$. In this case, $\omega_{jn}^{-1} \sim 0.1\ fs$. The magnetic field lifetime ($\sim 10$ picoseconds) is many orders of magnitude larger than a femtosecond, thus the magnetic field will exist sufficiently long enough so that the Weibler spectrum may be resolved. Furthermore, since the field-variability time-scale is $\sim$ picoseconds, which is much longer, the magnetic field may be treated as static.

\subsection{The Weibler Power}

Now, we will estimate the volumetric power of Weibler radiation to ascertain its observability using current instrumentation. We will ignore any magnetic anisotropy, statistical inhomogeneity, and plasma dispersion effects. We will consider a distribution of mono-energetic electrons that radiate isotropically. Since the characteristic wavelength of the emitted radiation by a single electron is smaller than the volume dimensions considered, we will assume that the radiation of the individual electrons add incoherently. Thus, with these assumptions, and the experimental values used previously, the volumetric radiated power is \citep{keenan15}
\begin{equation}
\frac{dP}{dV} \sim \frac{2}{3}n_e{(r_e\gamma_{min}\beta_{min}B_{max})}^2.
\label{jitter_power_def}
\end{equation} 
where $n_e$ is the number density of electrons in volume $dV$, and $r_e = e^2/m_e c^2$ is the classical electron radius. We expect the Weibel fields to predominantly reside at scales comparable to the electron skin-depth. Since the fields will, likely, be strongest at the site of laser absorption, i.e.\ the critical surface, we may very roughly estimate the Weibler power by substituting $n_e \sim n_c$ -- where the $n_c$ is the critical density, i.e.\
\begin{equation}
 n_c = \frac{m_e\omega^2}{4\pi e^2},
\label{crit_freq}
\end{equation} 
where $\omega$ is the laser frequency. Thus, we estimate the volumetric Weibler power as:
\begin{equation}
\frac{dP^{Weibler}}{dV} \sim 10^{22} \ erg \ {cm^{-3}} \ {s^{-1}}.
\label{jitter_power}
\end{equation} 
Finally, we should compare this result to estimates for any competing radiation mechanisms. We believe that \emph{thermal Bremssrahlung} (Bremss.) due to electron-ion collisions is the only likely complication. In the next subsection, we will make an attempt to roughly estimate the Bremss.\ contribution.

\section{Thermal Bremssrahlung}
\label{s:brems}

To estimate the electron-ion Bremss.\ component, we will assume a thermal distribution of electrons. We will assume, as before, the estimate for the  ``effective'' electron temperature, i.e.\ $T_e = 300-600\ keV$, obtained from the Ref. \citep{mondal12} PIC simulations. At these temperatures, the aluminum coating layer will be fully ionized, meaning $Z = 13$. Ignoring the particle escape from the aluminum layer (either into the vacuum or the BK-7 glass), the number density of ions $n_i = n_e/Z$. Thus, neglecting self-absorption (which only occurs at small frequencies), the electron Bremss.\ power per unit volume (in cgs units) will be \citep{rybicki}
\begin{equation}
\frac{dP}{dV} \sim 1.4\times10^{-27}{T_e}^{1/2}{n_e}{n_i}{Z}^2.
\label{bremss_pow_vol}
\end{equation} 
Now, we suspect that Bremss.\ radiation will be emitted throughout the entirety of the plasma. Nonetheless, owing to the square dependence on the plasma density, the regions of high-density will dominate the total emission power. 

Thus, we need an estimate of the density profile. To that end, we adopt the electron density supposed by the Ref. \citep{mondal12} PIC simulations. This was an exponential profile, in the longitudinal direction, of the form:
\begin{equation}
n_e(z) = \text{exp}(z/L - 1),
\label{num_dens}
\end{equation} 
where $L = 2\lambda$ is the scale length, and $z$ is the longitudinal coordinate. The profile was uniform in the transverse plane. This longitudinal trend continued up to a plateau at $n_e = 140n_c$. Then, the simulation box ended at $z = 16\lambda$. We adopt this profile here.

Finally, for our estimate of the Bremss.\ component, we will suppose that $n_e = 140n_c$. With this substitution, we have:
\begin{equation}
\frac{dP^{Bremss.}}{dV} \sim 10^{26} \ erg \ {cm^{-3}} \ {s^{-1}}.
\label{bremss_pow}
\end{equation} 
This value is four orders of magnitude larger than the Weibler radiation power. However, this estimate does not account for the variation in the power across the frequency domain. For this, we will need to estimate the radiation spectrum. As we will show, the Weibler spectrum dominates at low frequencies.

As a final consideration, we must ensure that these radiative processes are not obscured by the inevitable loss of particle energy via \emph{radiative cooling}. This requires that we estimate the cooling time-scales.

\section{Radiative Cooling}
\label{s:rad_cool}

First, we consider the Bremss.\ cooling time. Considering the electrons as a classical mono-atomic gas, the Bremss.\ cooling time, with $n_e = 140n_c$, is
\begin{equation}
t_{cool}^{Bremss.} \sim \frac{3n_ek_BT_e}{\left(\frac{dP^{Bremss.}}{dV}\right)} \sim 100 \ \mu{s},
\label{bremss_cool}
\end{equation}
which is a few orders of magnitude larger than all other time-scales in this experiment. Thus, Bremss.\ cooling is negligible.

The Weibler cooling time-scale may be estimated by considering the time at which the radiated power, for a given electron, is comparable to that electron's initial kinetic energy, i.e.\
\begin{equation}
P^{Weibler}_\text{single}{t^{Weibler}_{cool}} \sim (\gamma_e-1){m_e}c^2,
\label{jitter_cool}
\end{equation}
where $P^{Weibler}_\text{single}$ is the power emitted by a single particle -- i.e.\ Eq.\ (\ref{jitter_power_def}) divided by $n_e$. Using $\gamma_e \approx 1.59$, the Weibler cooling time is 
\begin{equation}
t^{Weibler}_{cool} \sim 0.1 \ \mu{s},
\label{jitter_cool_num}
\end{equation}
which is, also, sufficiently long enough to be ignored. We may conclude that, neither Bremss.\, nor Weibler cooling, is significant.

\section{The Radiation Power Spectrum}
\label{s:pow_spec}

Finally, we make predictions for the spectral profile of the emitted radiation. We retain our initial assumptions that the magnetic turbulence is statistically isotropic, that the electron density has the exponential (longitudinal) profile -- Eq.\ (\ref{num_dens}) -- that plateaus at $n_e = 140n_c$, and that the electron velocities are thermally distributed. For simplicity, we will assume an isotropic three-dimensional magnetic turbulence with a power-law turbulent spectrum:
\begin{equation}
\left\{\begin{array}{ll}
|{\bf B}_{\bf k}|^2 = Ck^{-\mu}, & k_{min} \le k \le k_{max}, 
 \\
|{\bf B}_{\bf k}|^2 = 0, & \text{otherwise}.
\end{array}\right.
\label{Bk}
\end{equation} 
Here the magnetic spectral index, $\mu$ is a free parameter, and
\begin{equation}
C \equiv \frac{2\pi^2V\langle B^2 \rangle}{\int_{k_\text{min}}^{k_\text{max}} \! k^{-\mu+2}\, \mathrm{d}k},
\label{C_def}
\end{equation} 
is a normalization, such that 
\begin{equation}
V^{-1}\int \! {\bf B}^2({\bf x}) \mathrm{d}{\bf x} = (2\pi)^{-3}\int \! |{\bf B}_{\bf k}|^2 \, \mathrm{d} {\bf k},
\label{C_def_cont}  
\end{equation} 
where $V$ is the volume of the space under consideration, and $\langle B^2 \rangle$ is the spatial average of the (mean free) magnetic field. 

Next, the thermal Bremss.\ power spectral density (i.e.,\ radiated power per frequency per unit volume) is a well known function:
\begin{equation}
\frac{dP}{d\omega{dV}} = \frac{8\sqrt{2}}{3\sqrt{\pi}}\sqrt{\epsilon(\omega)}\left[Z^2n_in_er_e^3\right]\frac{(m_ec^2)^{3/2}}{(k_BT_e)^{1/2}}\bar{G}(\omega, T_e),
\label{bremss_spec}
\end{equation}
where $\sqrt{\epsilon(\omega)}$ is the scalar dielectric permittivity, and $\bar{G}(\omega, T_e)$ is the velocity-averaged Gaunt factor. For high-temperature, though non-relativistic, electrons: \citep{bekefi}
\begin{equation}
\bar{G}(\omega, T_e) = \ln\left(\frac{4}{\gamma}\frac{k_BT_e}{\hbar\omega}\right),
\label{gaunt_fac}
\end{equation}
where $\gamma \approx 0.5772$ is the Euler–Mascheroni constant. Since the electron velocities are only mildly relativistic, the relativistic correction to Eq. (\ref{gaunt_fac}) will be relatively small -- a factor of a few. 

We may obtain the total Bremss.\ spectral flux by integrating Eq.\ (\ref{bremss_spec}) over the length of $z$, i.e.\
\begin{equation}
\frac{dP}{d\omega{dA}} = \frac{8\sqrt{2}}{3\sqrt{\pi}}\frac{(m_ec^2)^{3/2}}{(k_BT_e)^{1/2}}\bar{G}(\omega, T_e)Z^2r_e^3 \int \! \sqrt{\epsilon(\omega(z))}n_i(z)n_e(z) \, \mathrm{d}z,
\label{bremss_spec_flux}
\end{equation}
where $dA$ is the differential cross-section, and $dP$ is the differential radiant power.

Next, the non-relativistic Weibler spectrum for a single electron moving through statistically homogeneous/isotropic, static, sub-Larmor-scale magnetic turbulence has been derived previously (see Ref.\ \citep{keenan15} for details). Repeating the results, for the spectral distribution given by Eq.\ (\ref{Bk}), we have:
\begin{equation}
\frac{dP}{d\omega} \propto 
\left\{\begin{array}{ll}
A + D\omega^2, &\text{if}~ \omega \leq \omega_\textrm{jn}
 \\
F\omega^{-\mu+2} + G\omega^2 + K, &\text{if}~ \leq \omega_\textrm{bn} 
 \\
0, &\text{if}~  \omega > \omega_\textrm{bn},
\end{array}\right.
\label{analy_spec}  
\end{equation} 
where, if $\mu \neq 2$:
\begin{equation}
A \equiv \frac{v}{2-\mu}\left(k_\text{max}^{-\mu+2}-k_{min}^{-\mu+2}\right),
\label{A_def}  
\end{equation} 
\begin{equation}
 D \equiv -\frac{1}{v\mu}\left(k_\text{max}^{-\mu}-k_\text{min}^{-\mu}\right),
\label{D_def}  
\end{equation} 
\begin{equation}
 F \equiv \frac{v^\mu}{v}\left(\frac{1}{\mu-2} + \frac{1}{\mu}\right),
\label{F_def}  
\end{equation} 
\begin{equation}
 G \equiv  - \frac{1}{v\mu}k_\text{max}^{-\mu},
\label{G_def}  
\end{equation} 
\begin{equation}
 K \equiv \frac{v}{2-\mu}k_\text{max}^{-\mu+2}.
\label{K_def}  
\end{equation} 
The shape of this spectrum is relatively similar in the mildly relativistic regime; which we access by a Lorentz transformation on Eq.\ (\ref{analy_spec}). Notice that the spectral shape depends upon the magnetic spectral index, $\mu$. Furthermore, the spectrum peaks at the Weibler frequency, $\omega_\text{jn}$ -- thus, one may readily obtain the largest spatial scale of the magnetic turbulence directly from the Weibler spectrum.

Next, to obtain the total Weibler spectrum from a thermal distribution of electrons, we must average the single electron spectrum over the appropriate Maxwell-Boltzmann distribution, i.e.\
\begin{equation}
\frac{dP}{d\omega{dV}} = n_e\frac{\int \!  \ P_j(\omega, \omega_\text{pe}, p)e^{\sigma(1-\gamma_e)} \, \mathrm{d}^3p}{\int \!  \ e^{\sigma(1-\gamma_e)} \, \mathrm{d}^3p},
\label{jitter_spec}
\end{equation}
where $\sigma \equiv m_ec^2/k_BT_e$, and $P_j(\omega, \omega_\text{pe}, p) \equiv \frac{dP}{d\omega}(p, \omega_\text{pe})$ is the single electron spectrum with kinetic momentum $p = \gamma_em_ev$. To account for the mildly relativistic velocities, we have made the substitution for the usual variables: $1/2v^2 \rightarrow (\gamma_e - 1)$ and $p \rightarrow \gamma_ep$.

Thus, assuming the Weibler prescription for the entirety of the plasma length, the spectral flux of Weibler radiation will be:
\begin{equation}
\frac{dP}{d\omega{dA}} = \int \!  n_e(z)\frac{\int \!  \ P_j(\omega, \omega_\text{pe}(z), p)e^{\sigma(1-\gamma_e)} \, \mathrm{d}^3p}{\int \!  \ e^{\sigma(1-\gamma_e)} \, \mathrm{d}^3p} \, \mathrm{d}z,
\label{jitter_spec_flux}
\end{equation}

Due to non-perturbative effects, the low-frequency end of the Weibler radiation spectrum will differ slightly from the Weibler prescription by the addition of an $\omega^{1/2}$ power-law asymptote (see Ref.\ \citep{teraki14} for a detailed description). Given $\rho \sim 4$ and $\omega_{pe} \sim \omega_{jn}$, this deviation will be present near $\omega_{pe}$; it has no effect, however, on the high-frequency end of the spectrum. Consequently, we have elected to ignore this feature. 

As stated previously, a cyclotron/synchrotron component, corresponding to the mean magnetic field, will be present. However, since this component is largely screened out by plasma dispersion, and its effect is already well known, we omit it here. 

Additionally, we safely ignore the damping effect of Coulomb collisions, since the experimental Reynold's number is $Re_{exp} \sim \omega_{pe}/\nu_{ei} \sim 10^6$ -- where $\nu_{ei}$ is the electron-ion collision frequency \citep{mondal12}. From this, we may infer that $\omega_{jn} \gg \nu_{ei}$. 

Finally, we neglect the plasma gyrotropy. Since $\omega_{ce} < \omega_{pe}$, the gyrotropy will not be critically important to the plasma dispersion at high frequencies, i.e.\ near $\omega_{bn}$. 

Thus, we consider an isotropic, collisionless plasma. The scalar dielectric permittivity is, consequently
\begin{equation}
\epsilon(\omega) = 1 - \omega_{pe}^2/\omega^2.
\label{eps_def}
\end{equation}

Finally, we may construct the radiation power spectrum. In each plot, the relevant parameters are: $\mu = 4$, $\omega_{pe} = 3\times10^{15} \ rad/s$, $k_{min} = 0.5\omega_{pe}/c$, $k_{max} = 10k_{min}$, $n_e = 3\times10^{21} \ cm^{-3}$, $n_i = n_e/Z$, $k_BT_e = 300 \ keV$, and $\langle B^2 \rangle^{1/2} = 63 \ MG$. The Weibler spectrum was constructed using a logarithmically spaced, discretized range of electron velocities from $\beta_{min} = 0.1$ to $\beta_{max} = 0.99$.

\begin{figure}
\includegraphics[angle = 0, width = 1\columnwidth]{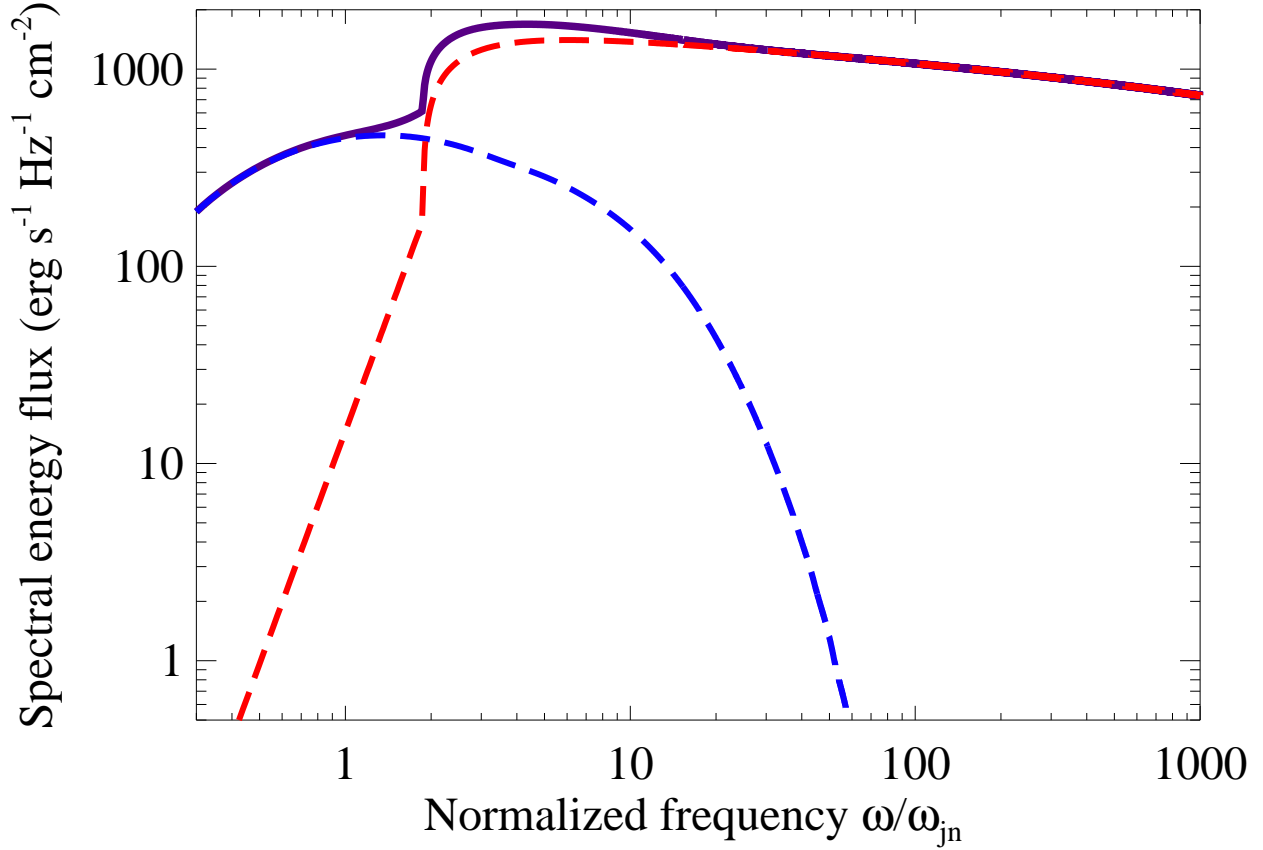}
\caption{(Color online) Spectral flux (differential flux per differential frequency) of the total emitted radiation vs normalized frequency. The frequency is normalized to the Weibler frequency, i.e.\ Eq.\ (\ref{jitter_freq}). The total power (``purple'', solid line) is the sum of the individual Weibler (``blue'', upper-left dashed line) and Bremssrahlung (``red'', lower-left dashed line) components. Clearly, the Weibler component dominants near the Weibler frequency (here defined as $f = \omega/2\pi$). }
\label{energy_flux}
\end{figure}

In Figure \ref{energy_flux}, the total spectral flux is plotted (``purple'', solid line) alongside the individual thermal Bremssrahlung (``red'', lower-left dashed line) and Weibler (``blue'', upper-left dashed line) components. Notice that the Weibler component dominants at frequencies near the Weibler frequency, $\omega_\text{jn} \sim \gamma_e^2k_\text{min}\beta c$ -- where $1/\sqrt{1 - \beta^2} = \gamma_e$ and $(\gamma_e-1)m_ec^2 \sim k_BT_e$. 

\begin{figure}
\includegraphics[angle = 0, width = 1\columnwidth]{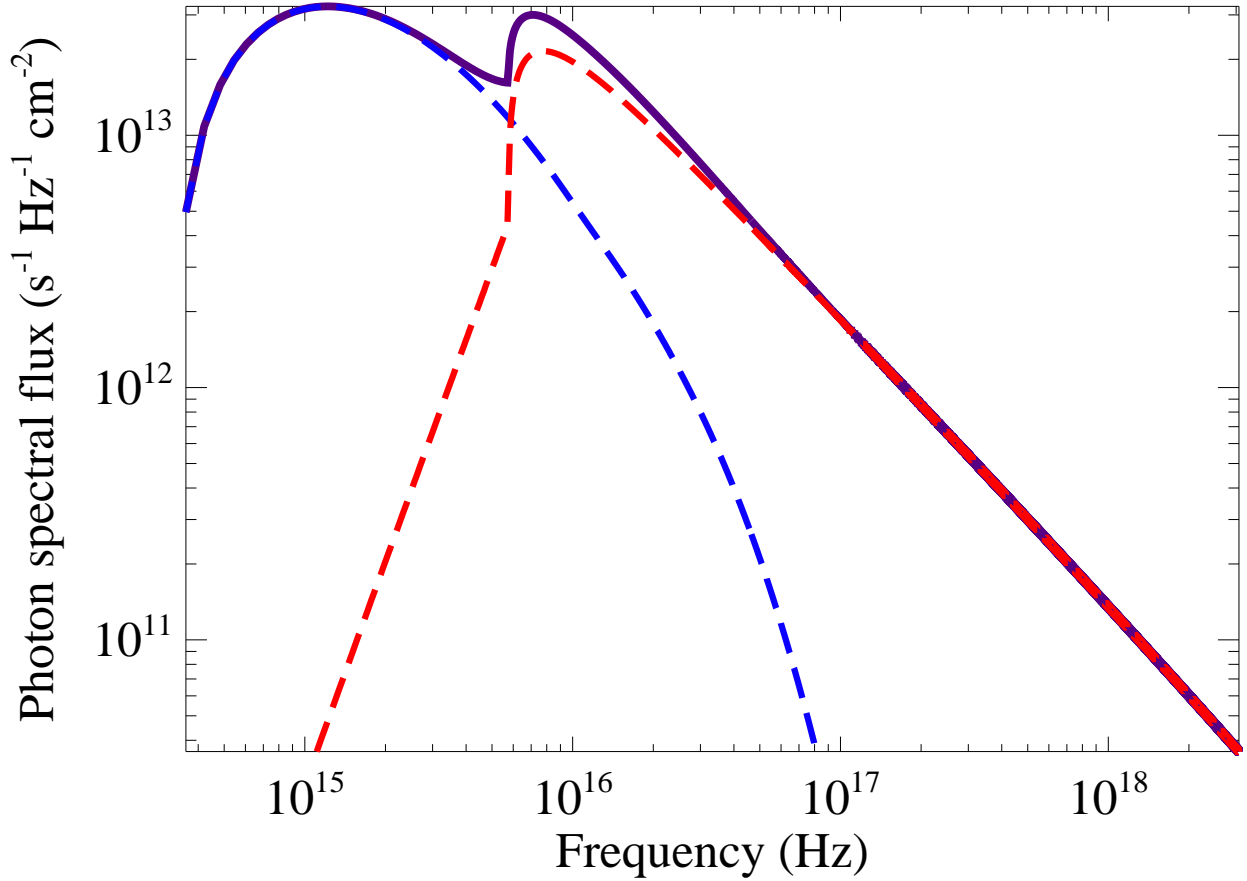}
\caption{(Color online) The photon flux at each frequency; i.e.\ Figure \ref{energy_flux} divided by the the photon energy, $\hbar\omega$. We see that the majority of the Weibler flux is at frequencies slightly below the Weibler frequency, $\omega_\text{jn}$.}
\label{photon_flux}
\end{figure}

Next, Figure \ref{photon_flux} displays the photon flux at each frequency; i.e.,\ Figure \ref{energy_flux} divided by the the photon energy, $\hbar\omega$. By integrating these curves over the complete frequency range, we may estimate the total photon flux for each component. These are $2\times10^{29} \ (photons) \ cm^{-2} \ s^{-1}$ and $10^{30} \ (photons) \ cm^{-2} \ s^{-1}$ for Weibler and Bremss., respectively. Thus, it would appear that the Bremss.\ flux is only an order of magnitude larger than the Weibler flux.
Since Bremss.\ emission is easily and routinely detectable in plasma experiments, it should be easy to observe Weibler radiation too. It is the very distinct spectral shapes of the two, along with the high fluxes, that make Bremss.\ and Weibler radiation easily distinguishable form one another and allows one to resolve their spectral features well -- the key factor of a good plasma diagnostic tool.

\section{Conclusions}
\label{s:concl}

In this paper we have investigated the prospects for the direct radiative diagnostics of a mildly relativistic, solid-density laser plasmas produced in current lab experiments. Our results demonstrate the feasibility of such an approach. Particularly, our analysis shows that a sub-relativistic laser-plasma setup, such as the experiment described in Mondal, et al. \citep{mondal12}, is a promising candidate for the direct observation of mildly relativistic Weibler radiation. We believe this is sufficient impetus for experimental exploration. 

Our model is, nonetheless, a simplification. To produce results, we had to make a number of assumptions. The most important of these concerns isotropy -- both in the magnetic turbulence and the emission of radiation. The turbulent magnetic fields produced by Weibel-like instabilities are typically characterized by anisotropy. This is because the distribution function of particles that produce Weibel fields is, itself, anisotropic. Thus, our assumption that an isotropic Maxwell-Boltzmann distribution, with a given ``effective'' temperature for the entire plasma, is not likely to hold on initial time-scales.

Similarly, our assumption that the magnetic turbulence is statistically homogeneous -- i.e.\ characterized by a single spectral distribution throughout the plasma -- is suspect. The correlation length throughout the plasma is likely, itself, a function of the electron density. For this reason, there may be regions within the solid target where the magnetic field is not sub-Larmor-scale; hence, the small-angle Weibler prescription fails there.

Nevertheless, we believe our model is reasonable. Our model illustrates two key features that we expect will be present in real lab experiments. First, the Weibler spectrum peaks near the frequency, $\omega_\text{jn} = \gamma_e^2k_\text{min}\beta c$, where $k_\text{min}$ is the characteristic wave-number of the magnetic turbulence. Thus, we may directly extract the correlation length, $\lambda_B \sim k_\text{min}^{-1}$, from the radiation spectrum. Lastly, the Weibler spectrum takes a sharp drop near the second break, $\omega_\text{bn} = \gamma_e^2k_\text{max}\beta{c}$. Similarly, $k_\text{max}$ denotes the minimum spatial scale. Although this feature may be concealed by the Bremss.\ component, we may extract it by subtracting the predicted Bremss.\ spectrum.

A very important feature of our model, as an advanced radiative diagnostic tool, is the ability to probe plasmas at different locations (depths). Indeed, Bremss.\ is a quadratic function of the density, so this radiation probes the plasma conditions in the densest parts of the plasma, i.e., deep into the ``core''. In contrast, Weibler radiation probes the region with the strongest small-scale fields, which occur where the laser energy/momentum deposition is most efficient, i.e., near the critical surface. The location of this region depends on both plasma density and the laser frequency, which opens up a possibility to do some sort of ``plasma tomography'' by using different laser frequencies. 

To conclude, we believe these preliminary results provide sufficient impetus for experimental exploration. If Weibler radiation is observed, it will provide valuable information about the statistical properties of the underlying small-scale, magnetic turbulence.

\begin{acknowledgments}
This work has been supported by the DOE grant DE-FG02-07ER54940 and the NSF grant AST-1209665.
\end{acknowledgments}

\end{document}